\documentclass[11pt,a4paper]{article}
\usepackage[T1]{fontenc}
\usepackage[utf8]{inputenc}
\usepackage{lmodern}
\usepackage[margin=2.5cm]{geometry}
\usepackage{amsmath,amssymb,bm}
\usepackage{graphicx}
\usepackage{authblk}
\usepackage{microtype}
\usepackage[hidelinks,hypertexnames=false]{hyperref}
\newcommand{\aGB}{\alpha_{\rm GB}}
\newcommand{\lp}{\ell_{P,5}}
\makeatletter
\newcommand{\clearsqrt}[1]{\mathpalette\clear@sqrt{#1}}
\newcommand{\clear@sqrt}[2]{#1\sqrt{%
 \vphantom{\raisebox{2.5pt}{$\m@th#1 #2$}}#2}}
\makeatother

\title{Extended Generalized Uncertainty Principle Corrections to
Five-Dimensional Einstein--Gauss--Bonnet Black-Hole Thermodynamics}
\author[1]{Erdin\c{c} Ula\c{s} Saka\thanks{Electronic address:
\href{mailto:ulassaka@istanbul.edu.tr}{\texttt{ulassaka@istanbul.edu.tr}}}}
\author[2]{Bekir Can L\"utf\"uo\u{g}lu}
\affil[1]{Department of Physics, Faculty of Science,
Istanbul University, Istanbul, T\"urkiye}
\affil[2]{Department of Physics, Faculty of Science,
University of Hradec Kr\'{a}lov\'{e}, Hradec Kr\'{a}lov\'{e}, Czechia}

\date{}
\begin{document}
\maketitle

\begin{abstract}
We investigate the thermodynamics of an asymptotically flat
five-dimensional Einstein--Gauss--Bonnet black hole within the extended
generalized uncertainty principle. A geometry-dependent localization
scale is fixed by requiring the undeformed result to reproduce the
surface-gravity temperature. The resulting expression has the correct
GUP, EUP, and semiclassical limits and imposes a joint constraint on the
minimum-length and minimum-momentum parameters. Because the effective
localization scale is nonmonotonic, the reality condition does not
generically yield a single minimum horizon radius. Depending on the
parameters, the admissible domain remains connected, develops a
degenerate critical radius, or separates into two disconnected branches.
We derive the corresponding entropy, heat capacity, and Helmholtz free
energy and examine their behavior numerically. The Gauss--Bonnet coupling
shifts and suppresses the near-horizon temperature maximum, while the EUP
sector produces a temperature minimum followed by linear growth at large
radius. In the two-branch regime, the outer endpoint has finite
temperature and vanishing heat capacity but is approached from the
locally unstable side. It should therefore be regarded as a limiting
thermodynamic configuration rather than a dynamically established
remnant. We also show that the perturbative entropy departs from the exact
result at large radius and that the leading GUP entropy correction
survives in the Einstein-gravity limit.
\end{abstract}

\begin{center}\small\textit{Keywords:} Extended generalized uncertainty principle; Einstein--Gauss--Bonnet gravity; black-hole thermodynamics; thermodynamic stability.\end{center}

\section{Introduction}
\label{sec:introduction}

Einstein--Gauss--Bonnet {(EGB)} gravity provides one of the simplest settings in
which higher-curvature corrections modify black-hole physics while the
field equations remain second order. The Gauss--Bonnet {(GB)} combination
constitutes the quadratic member of the Lovelock hierarchy
\cite{Lovelock1971} and also arises as a leading higher-curvature
correction in low-energy effective string theory. The corresponding field
equations admit static black-hole solutions with two branches
\cite{BoulwareDeser1985}, of which only the minus branch possesses a smooth
Einstein-gravity limit. The thermodynamic interpretation rests on the
Bekenstein--Hawking relation between horizon geometry, entropy, and
temperature \cite{Bekenstein1973,Hawking1975}. In higher-curvature gravity,
however, the entropy is no longer given by the area term alone: the
{GB} coupling enters both the temperature and the Wald entropy
\cite{MyersSimon1988,Wald1993,Cai2002}.

Close to the Planck scale, several approaches to quantum gravity suggest
that the standard Heisenberg uncertainty relation should be modified by
the appearance of a minimum resolvable length
\cite{Hossenfelder2013}. A widely studied realization is the generalized
uncertainty principle (GUP), which introduces this scale through a
momentum-dependent correction
\cite{Maggiore1993,KempfManganoMann1995}.  When applied to black-hole
thermodynamics, the GUP modifies the Hawking temperature and can replace
the usual divergent endpoint of evaporation with a finite limiting mass
\cite{AdlerChenSantiago2001,MedvedVagenas2004}.  Such a limiting configuration is often
described as a black-hole remnant.  However, the existence of a minimum
radius or a vanishing heat capacity does not by itself establish that
evaporation dynamically stops.

An infrared modification can be introduced through the extended
uncertainty principle (EUP).  In contrast to the GUP correction, which
becomes important at short distances, the EUP contribution grows with
the position uncertainty and produces a minimum momentum.  It appears
naturally in discussions of uncertainty relations in curved backgrounds
and black-hole thermodynamics
\cite{BolenCavaglia2005,Park2008,Mignemi2010}.  Combining the ultraviolet and
infrared corrections leads to the extended generalized uncertainty
principle (EGUP), which contains both a minimum-length and a
minimum-momentum sector.  In the present asymptotically flat setting, the
large-distance scale entering the EUP term is treated as an independent
infrared deformation scale; it is not identified with a cosmological
curvature radius.

The thermodynamic consequences of generalized uncertainty relations have
been investigated for a wide variety of black-hole geometries.  These
studies indicate that GUP corrections can modify the Hawking temperature,
entropy, heat capacity, and the endpoint structure of evaporation, while
EUP-type corrections become particularly relevant in the large-distance
regime.  Extended uncertainty relations have also been applied to charged
AdS black holes in {GB} gravity, together with analyses of their
thermodynamic and optical properties
\cite{Chaudhary:2021uuk}.  Such studies demonstrate that the
interplay between higher-curvature effects and uncertainty-principle
deformations can substantially alter the thermodynamic behavior.
Nevertheless, the resulting predictions depend sensitively on the
background geometry, the adopted uncertainty relation, and the
identification of the position uncertainty with a characteristic
black-hole length scale.

More recently, the GUP thermodynamics of a five-dimensional
{EGB} black hole has been considered in
Ref.~\cite{PradhanVargheseFairoos2026}.  That analysis focused on the
short-distance deformation associated with the GUP.  The simultaneous
inclusion of ultraviolet and infrared corrections in the asymptotically
flat five-dimensional geometry, however, requires a separate treatment.
In particular, the EGUP extension cannot be obtained simply by adding a
small infrared correction to the GUP result.  The two deformation sectors
are constrained jointly by the requirement that the uncertainty relation
admit real solutions.  Moreover, recovering the surface-gravity
temperature in the undeformed limit requires a geometry-dependent
calibration of the position uncertainty.  This calibration makes the
effective localization scale nonmonotonic in the horizon radius and
changes the structure of the admissible thermodynamic domain.

The purpose of this work is to develop the EGUP-modified thermodynamics
of an asymptotically flat five-dimensional {EGB} black
hole, with particular attention to these issues. We first derive the physical
five-dimensional {EGB} branch directly from the field
equations and keep separate the coupling in the action and the effective
coupling appearing in the metric.  We then determine the EGUP temperature
by requiring the undeformed uncertainty relation to reproduce the
surface-gravity result.  The GUP, EUP, and semiclassical temperatures are
recovered as independent limits. We show that the temperature reality condition can lead to three
qualitatively different situations: a connected positive-radius domain,
a degenerate critical radius, or two disconnected admissible branches.
The latter case contains an excluded interval that cannot be crossed by a
black hole evaporating along the large-radius branch.  We derive the
corresponding entropy, heat capacity, and Helmholtz free energy and examine
their behavior numerically.  The analysis also shows that the leading GUP
entropy correction remains nonzero in the Einstein-gravity limit, while
the EUP sector changes the large-radius temperature from inverse-radius
decay to linear growth.  Particular care is taken to distinguish a
thermodynamic limiting configuration from a dynamically stable remnant.

The paper is organized as follows.  In Sec.~\ref{sec:egb-black-hole}, we
derive the five-dimensional {EGB} black-hole solution
and summarize its semiclassical thermodynamics.  In
Sec.~\ref{sec:egup-temperature}, we construct the EGUP-modified
temperature and analyze its reality condition.  The corrected entropy,
heat capacity, and Helmholtz free energy are obtained in
Sec.~\ref{sec:thermodynamics}.  The dimensionless critical structure is
developed in Sec.~\ref{sec:critical-behavior}, and the numerical results
are presented in Sec.~\ref{sec:numerical}.  Our conclusions are collected
in Sec.~\ref{sec:conclusion}.

\section{Five-dimensional \texorpdfstring{{EGB}}{EGB} black hole}
\label{sec:egb-black-hole}

We begin with the five-dimensional {EGB} action in vacuum,
\begin{equation}
 I=\frac{1}{16\pi G_5}\int d^5x\sqrt{-g}\left(
 R+\alpha_{\mathrm{action}}\mathcal{L}_{\mathrm{GB}}\right),
 \label{eq:action}
\end{equation}
where
\begin{equation}
 \mathcal{L}_{\mathrm{GB}}
 =R^2-4R_{\mu\nu}R^{\mu\nu}
 +R_{\mu\nu\rho\sigma}R^{\mu\nu\rho\sigma}.
 \label{eq:gb-density}
\end{equation}
Varying Eq.~\eqref{eq:action} with respect to the metric gives
\begin{equation}
 G_{\mu\nu}+\alpha_{\mathrm{action}}H_{\mu\nu}=0,
 \label{eq:field-equations}
\end{equation}
with
\begin{align}
 H_{\mu\nu}={}&2\left(
 RR_{\mu\nu}-2R_{\mu\lambda}R^{\lambda}{}_{\nu}
 -2R^{\lambda\sigma}R_{\mu\lambda\nu\sigma}
 +R_{\mu}{}^{\lambda\rho\sigma}R_{\nu\lambda\rho\sigma}
 \right)
 -{\frac{1}{2}}g_{\mu\nu}\mathcal{L}_{\mathrm{GB}}.
 \label{eq:lanczos}
\end{align}

For a static and spherically symmetric spacetime, we take
\begin{equation}
 ds^2=-f(r)dt^2+\frac{dr^2}{f(r)}+r^2d\Omega_3^2,
 \label{eq:metric-ansatz}
\end{equation}
where $d\Omega_3^2$ denotes the line element of the unit three-sphere,
whose area is $\Omega_3=2\pi^2$.  It is useful to introduce the effective
{GB} parameter
\begin{equation}
 \aGB=(D-3)(D-4)\alpha_{\mathrm{action}}=2\alpha_{\mathrm{action}}
 \qquad (D=5),
 \label{eq:effective-alpha}
\end{equation}
so that no ambiguity arises between the coupling in the action and the
parameter appearing in the metric.  With the ansatz~\eqref{eq:metric-ansatz},
the independent radial field equation can be written as
\begin{equation}
 \frac{d}{dr}\left[
 r^2\bigl(1-f(r)\bigr)
 +\aGB\bigl(1-f(r)\bigr)^2
 \right]=0.
 \label{eq:reduced-field-equation}
\end{equation}
After one integration, Eq.~\eqref{eq:reduced-field-equation} becomes
\begin{equation}
 r^2\bigl(1-f(r)\bigr)
 +\aGB\bigl(1-f(r)\bigr)^2=\mu,
 \label{eq:integrated-field-equation}
\end{equation}
where the integration constant is related to the ADM mass by
\begin{equation}
 \mu=\frac{16\pi G_5M}{(D-2)\Omega_3}
 =\frac{8G_5M}{3\pi}.
 \label{eq:mass-parameter}
\end{equation}
Solving the quadratic equation~\eqref{eq:integrated-field-equation} gives
two branches,
\begin{equation}
 f_{\pm}(r)=1+\frac{r^2}{2\aGB}
 \left[1\pm\sqrt{1+\frac{4\aGB\mu}{r^4}}\right].
 \label{eq:two-branches}
\end{equation}
The minus branch has a smooth general-relativistic limit,
$f_-(r)\to1-\mu/r^2$ as $\aGB\to0$, whereas the plus branch does not.
We therefore work exclusively with the asymptotically flat minus branch,
\begin{equation}
 f(r)=1+\frac{r^2}{2\aGB}
 \left[1-\sqrt{1+\frac{32\aGB G_5 M}{3\pi r^4}}\right],
 \label{eq:metric-function}
\end{equation}
in units $c=\hbar=k_B=1$.  The horizon equation gives
\begin{equation}
 M(r_+)=\frac{3\pi}{8G_5}\left(r_+^2+\aGB\right).
 \label{eq:mass-horizon}
\end{equation}
The surface-gravity temperature and the Wald entropy are
\begin{align}
 T_H(r_+)&=\frac{r_+}{2\pi(r_+^2+2\aGB)},\\
 S_0(r_+)&=\frac{A_3}{4G_5}
 \left(1+\frac{6\aGB}{r_+^2}\right)
 =\frac{\pi^2}{2G_5}\left(r_+^3+6\aGB r_+\right),
\end{align}
where $A_3=2\pi^2r_+^3$.  In five dimensions the Planck length satisfies
\begin{equation}
 \lp^3=G_5{,}
\end{equation}
in natural units (and $\lp^3=\hbar G_5/c^3$ when constants are restored).
Thus the entropy scales as $A_3/\lp^3$, rather than $A_3/\lp^2$.

For later comparison, the semiclassical heat capacity is
\begin{equation}
 C_0=\frac{dM/dr_+}{dT_H/dr_+}
 =\frac{3\pi^2r_+(r_+^2+2\aGB)^2}
 {2G_5(2\aGB-r_+^2)}.
\end{equation}
It diverges at $r_+=\sqrt{2\aGB}$.

\section{EGUP-modified temperature}
\label{sec:egup-temperature}

We adopt the quadratic EGUP
\begin{equation}
 \Delta x\,\Delta p\geq\frac{1}{2}
 \left[1+\beta^2\lp^2(\Delta p)^2
 +\eta^2\frac{(\Delta x)^2}{L^2}\right],
\end{equation}
where $\beta$ and $\eta$ are dimensionless parameters and $L$ is the
large-distance scale associated with the EUP sector. {We then define}
\begin{equation}
 X(r_+)=r_++\frac{2\aGB}{r_+},\qquad
 \delta=\frac{\pi^2\eta^2}{L^2}.
\end{equation}
The calibration
\begin{equation}
 \Delta x=\pi X(r_+),\qquad \Delta p=T
 \label{eq:localization-calibration}
\end{equation}
reproduces the surface-gravity temperature when both deformation parameters
vanish.
The EGB background, uncertainty-based temperature prescription, and
geometry-dependent calibration are adopted from the GUP treatment of
Ref.~\cite{PradhanVargheseFairoos2026}, after identifying its metric
coupling with our $\alpha_{\rm GB}$. Here we include the EUP term and
examine its consequences for the joint reality condition and admissible
radius domains. The calibration is a phenomenological matching
prescription, not a derivation from a microscopic quantum-gravity theory.
Retaining it unchanged at nonzero deformation is a modeling assumption.
All results below are conditional on this prescription, and ``exact''
refers to the unexpanded expressions within this model.

Saturating the EGUP gives
\begin{equation}
 X=\frac{1}{2\pi}
 \left[\frac{1+\delta X^2}{T_{\rm EGUP}}
 +\beta^2\lp^2T_{\rm EGUP}\right].
\end{equation}
The branch possessing the correct undeformed limit is
\begin{equation}
 {
 T_{\rm EGUP}=\frac{\pi X}{\beta^2\lp^2}
 \left[1-\sqrt{1-
 \frac{\beta^2\lp^2(1+\delta X^2)}{\pi^2X^2}}
 \right] }.
 \label{eq:temperature-exact}
\end{equation}
Its limiting cases are
\begin{align}
 T_{\rm GUP}&=\frac{\pi X}{\beta^2\lp^2}
 \left[1-\sqrt{1-\frac{\beta^2\lp^2}{\pi^2X^2}}\right]
 &&(\eta=0){,}\\
 T_{\rm EUP}&=\frac{1+\delta X^2}{2\pi X}
 &&(\beta=0){,}\\
 T_H&=\frac{1}{2\pi X}
 &&(\beta=\eta=0){.}
\end{align}

The reality condition is
\begin{equation}
 (\pi^2-\beta^2\lp^2\delta)X^2\geq\beta^2\lp^2,
 \qquad \pi^2>\beta^2\lp^2\delta.
\end{equation}
Writing
\begin{equation}
 X_c=\frac{\beta\lp}{\sqrt{\pi^2-\beta^2\lp^2\delta}},
\end{equation}
the boundary radii, when $X_c^2\geq8\aGB$, are
\begin{equation}
 r_c^{\pm}=\frac{1}{2}
 \left(X_c\pm\sqrt{X_c^2-8\aGB}\right).
\end{equation}
Consequently, the EGUP does not generically produce a single critical mass:
depending on the parameters, there can be no excluded interval, one limiting
radius, or two disconnected admissible radius branches.

The reality condition can also be understood directly from the
uncertainty relation. Regarding the saturated EGUP as a quadratic
equation for $\Delta p$, a real momentum uncertainty exists only if
\begin{equation}
 (\Delta x)^2
 \left(1-\frac{\beta^2\eta^2\ell_{P,5}^2}{L^2}\right)
 \geq \beta^2\ell_{P,5}^2.
\end{equation}
Consequently, the minimum measurable length is
\begin{equation}
 (\Delta x)_{\min}
 =
 \frac{\beta\ell_{P,5}}
 {\sqrt{1-\beta^2\eta^2\ell_{P,5}^2/L^2}}.
 \label{eq:minimal-length}
\end{equation}
Similarly, treating the EGUP as a quadratic equation for $\Delta x$
gives the minimum measurable momentum
\begin{equation}
 (\Delta p)_{\min}
 =
 \frac{\eta/L}
 {\sqrt{1-\beta^2\eta^2\ell_{P,5}^2/L^2}}.
 \label{eq:minimal-momentum}
\end{equation}
Both quantities are well defined only if
\begin{equation}
 \frac{\beta\eta\ell_{P,5}}{L}<1.
 \label{eq:egup-parameter-bound}
\end{equation}
Thus, the GUP and EUP sectors are not completely independent: their
deformation parameters must satisfy the bound in
Eq.~\eqref{eq:egup-parameter-bound}.

It is worth clarifying the calibration used in Eq.~\eqref{eq:localization-calibration}. If the
position uncertainty is initially written as
\begin{equation}
 \Delta x=\epsilon(r_+)r_+,
\end{equation}
the requirement that the undeformed uncertainty relation reproduce
the surface-gravity temperature fixes
\begin{equation}
 \epsilon(r_+)
 =
 \pi\left(1+\frac{2\alpha_{\rm GB}}{r_+^2}\right).
\end{equation}
Therefore,
\begin{equation}
 \Delta x
 =
 \pi\left(r_++\frac{2\alpha_{\rm GB}}{r_+}\right)
 =\pi X(r_+).
\end{equation}
This geometry-dependent calibration incorporates the {GB}
correction into the effective localization scale of a Hawking
quantum. It should be regarded as a thermodynamic calibration rather
than a direct identification of $\Delta x$ with the horizon radius.

For sufficiently small deformation parameters, the exact
temperature in Eq.~\eqref{eq:temperature-exact} can be expanded as
\begin{equation}
 T_{\rm EGUP}
 =
 \frac{1+\delta X^2}{2\pi X}
 +
 \frac{\beta^2\ell_{P,5}^2(1+\delta X^2)^2}
 {8\pi^3X^3}
 +\mathcal{O}(\beta^4).
 \label{eq:temperature-beta-expansion}
\end{equation}
Keeping terms linear in $\delta$ and quadratic in $\beta$ gives
\begin{align}
 T_{\rm EGUP}
 ={}&
 T_H
 +\frac{\delta X}{2\pi}
 +\frac{\beta^2\ell_{P,5}^2}{8\pi^3X^3}
 +\frac{\beta^2\ell_{P,5}^2\delta}{4\pi^3X}
 +\mathcal{O}
 \left(\delta^2,\beta^4,\beta^2\delta^2\right).
 \label{eq:temperature-double-expansion}
\end{align}
The second and third terms in Eq.~\eqref{eq:temperature-double-expansion}
represent the leading EUP and GUP corrections, respectively, whereas
the fourth term is the leading mixed EGUP contribution. This term
cannot be obtained by considering the GUP and EUP sectors separately.

The function $X(r_+)$ is not monotonic. It reaches its minimum at
\begin{equation}
 r_*=\sqrt{2\alpha_{\rm GB}},
 \qquad
 X_{\min}=2\sqrt{2\alpha_{\rm GB}}.
 \label{eq:X-minimum}
\end{equation}
This observation leads to three qualitatively different cases:
\begin{enumerate}
 \item If $X_c^2<8\alpha_{\rm GB}$, the EGUP temperature is real
 for every $r_+>0$, and no additional cutoff radius is generated.

 \item If $X_c^2=8\alpha_{\rm GB}$, the two boundary radii coincide
 at $r_*= \sqrt{2\alpha_{\rm GB}}$.

 \item If $X_c^2>8\alpha_{\rm GB}$, the admissible domain splits into
 two disconnected branches,
 \begin{equation}
 0<r_+\leq r_c^-,
 \qquad\text{and}\qquad
 r_+\geq r_c^+,
 \end{equation}
 while the interval $r_c^-<r_+<r_c^+$ is excluded because the
 temperature becomes complex.
\end{enumerate}

The masses associated with the two boundary radii are
\begin{equation}
 M_c^\pm
 =
 \frac{3\pi}{8G_5}
 \left[
 \alpha_{\rm GB}+(r_c^\pm)^2
 \right].
 \label{eq:critical-masses}
\end{equation}
For an evaporating black hole evolving from the large-radius regime,
the physically accessible endpoint is $r_c^+$, and hence the relevant
limiting mass is $M_c^+$. The smaller solution $r_c^-$ belongs to a
disconnected branch and should not be interpreted as a second stage
of the same evaporation process.

At either boundary, the square root in Eq.~\eqref{eq:temperature-exact} vanishes and the
temperature takes the finite value
\begin{equation}
 T_c^\pm
 =
 \frac{\pi X_c}{\beta^2\ell_{P,5}^2}.
 \label{eq:critical-temperature}
\end{equation}
Whether this limiting configuration represents a dynamically stable
black-hole remnant cannot be inferred from the reality condition
alone. Its thermodynamic stability must be established independently
from the heat capacity, which will be analyzed in the next section.

\section{EGUP-corrected thermodynamic quantities}
\label{sec:thermodynamics}

Having obtained the EGUP-modified temperature, we now investigate
the corresponding entropy, heat capacity, and Helmholtz free energy.
For convenience, we introduce
\begin{equation}
 b=\beta^2\ell_{P,5}^2,
 \qquad
 Z(r_+)=
 \sqrt{
 1-\frac{b}{\pi^2}
 \left(
 \frac{1}{X^2}+\delta
 \right)
 },
 \label{eq:Z-definition}
\end{equation}
where
\begin{equation}
 X(r_+)=r_++\frac{2\alpha_{\rm GB}}{r_+},
 \qquad
 X'(r_+)=1-\frac{2\alpha_{\rm GB}}{r_+^2}.
 \label{eq:X-derivative}
\end{equation}
In terms of $Z$, the EGUP temperature can equivalently be written as
\begin{equation}
 T_{\rm EGUP}
 =
 \frac{1+\delta X^2}
 {\pi X(1+Z)}.
 \label{eq:temperature-rationalized}
\end{equation}
This rationalized form is particularly useful in the derivation of
the remaining thermodynamic quantities.

\subsection{EGUP-corrected entropy}
\label{subsec:entropy}

At fixed {GB} coupling, the first law takes the form
\begin{equation}
 dM=T_{\rm EGUP}\,dS_{\rm EGUP}.
 \label{eq:first-law}
\end{equation}
Using
\begin{equation}
 \frac{dM}{dr_+}=\frac{3\pi r_+}{4G_5},
\end{equation}
together with Eq.~\eqref{eq:temperature-rationalized}, we obtain
\begin{equation}
 \frac{dS_{\rm EGUP}}{dr_+}
 =
 \frac{3\pi^2 r_+X}{4G_5}
 \frac{1+Z}{1+\delta X^2}.
 \label{eq:entropy-derivative-exact}
\end{equation}
Therefore, the exact entropy can be represented by
\begin{equation}
 S_{\rm EGUP}(r_+)
 =
 \frac{3\pi^2}{4G_5}
 \int^{r_+}
 \frac{\bar r\,X(\bar r)
 \left[1+Z(\bar r)\right]}
 {1+\delta X^2(\bar r)}
 \,d\bar r+S_{\rm ref},
 \label{eq:entropy-integral}
\end{equation}
where $S_{\rm ref}$ is an integration constant whose value depends
on the thermodynamic reference state.

For small deformation parameters, the inverse temperature becomes
\begin{equation}
 \frac{1}{T_{\rm EGUP}}
 =
 \frac{2\pi X}{1+\delta X^2}
 -\frac{b}{2\pi X}
 +\mathcal{O}(b^2).
 \label{eq:inverse-temperature-expansion}
\end{equation}
Expanding also to first order in $\delta$, the entropy is
\begin{align}
 S_{\rm EGUP}
 ={}&
 S_0
 -\frac{3\pi^2\delta}{2G_5}
 \left(
 \frac{r_+^5}{5}
 +2\alpha_{\rm GB}r_+^3
 +12\alpha_{\rm GB}^2r_+
 -\frac{8\alpha_{\rm GB}^3}{r_+}
 \right)
 \nonumber\\
 &-\frac{3b}{8G_5}
 \left[
 r_+
 -\sqrt{2\alpha_{\rm GB}}\,
 \tan^{-1}
 \left(
 \frac{r_+}{\sqrt{2\alpha_{\rm GB}}}
 \right)
 \right]
 +S_{\rm ref}
 +\mathcal{O}
 \left(
 \delta^2,b^2
 \right),
 \label{eq:entropy-perturbative}
\end{align}
where
\begin{equation}
 S_0=
 \frac{\pi^2}{2G_5}
 \left(
 r_+^3+6\alpha_{\rm GB}r_+
 \right)
\end{equation}
is the standard Wald entropy.

The perturbative entropy requires
\begin{equation}
 \frac{b(1+\delta X^2)}{\pi^2X^2}\ll1,
 \qquad \delta X^2\ll1.
 \label{eq:entropy-validity}
\end{equation}
For an entropy difference, both conditions must hold throughout the
integration interval from the reference radius. For $\alpha_{\rm GB}>0$
and $\delta>0$, $X$ diverges as $r_+\to0^+$ and as $r_+\to\infty$,
so the EUP expansion fails in both limits. We use the exact entropy
integral outside the perturbative range.

Equation~\eqref{eq:entropy-perturbative} shows that the leading GUP
correction does not vanish in the limit $\alpha_{\rm GB}\to0$.
Indeed,
\begin{equation}
 \Delta S_{\rm GUP}\Big|_{\alpha_{\rm GB}\to0}
 =
 -\frac{3b}{8G_5}r_+.
 \label{eq:entropy-alpha-zero}
\end{equation}

\subsection{Heat capacity and local stability}
\label{subsec:heat-capacity}

The heat capacity is defined by
\begin{equation}
 C_{\rm EGUP}
 =
 \frac{dM}{dT_{\rm EGUP}}
 =
 \frac{dM/dr_+}{dT_{\rm EGUP}/dr_+}.
 \label{eq:heat-capacity-definition}
\end{equation}
Differentiating Eq.~\eqref{eq:Z-definition} gives
\begin{equation}
 \frac{dZ}{dr_+}
 =
 \frac{bX'}{\pi^2X^3Z}.
 \label{eq:Z-derivative}
\end{equation}
It follows from Eq.~\eqref{eq:temperature-rationalized} that
\begin{equation}
 \frac{1}{T_{\rm EGUP}}
 \frac{dT_{\rm EGUP}}{dr_+}
 =
 X'
 \left[
 \frac{2\delta X}{1+\delta X^2}
 -\frac{1}{X}
 -\frac{b}
 {\pi^2X^3Z(1+Z)}
 \right].
 \label{eq:temperature-log-derivative}
\end{equation}
Consequently, the exact heat capacity can be written compactly as
\begin{equation}
 {
 C_{\rm EGUP}
 =
 \frac{3\pi r_+}
 {4G_5T_{\rm EGUP}X'}
 \left[
 \frac{2\delta X}{1+\delta X^2}
 -\frac{1}{X}
 -\frac{b}
 {\pi^2X^3Z(1+Z)}
 \right]^{-1}
 }.
 \label{eq:heat-capacity-exact}
\end{equation}

The black hole is locally thermodynamically stable when
\begin{equation}
 C_{\rm EGUP}>0,
\end{equation}
and unstable when $C_{\rm EGUP}<0$. Divergences of the heat capacity
are determined by
\begin{equation}
 \frac{dT_{\rm EGUP}}{dr_+}=0,
 \label{eq:phase-transition-condition}
\end{equation}
and signal possible continuous thermodynamic phase transitions.

In the undeformed limit, Eq.~\eqref{eq:heat-capacity-exact} reduces to
\begin{equation}
 C_0
 =
 \frac{3\pi^2r_+
 \left(r_+^2+2\alpha_{\rm GB}\right)^2}
 {2G_5\left(2\alpha_{\rm GB}-r_+^2\right)},
 \label{eq:classical-heat-capacity-again}
\end{equation}
whose divergence occurs at
\begin{equation}
 r_+=\sqrt{2\alpha_{\rm GB}}.
\end{equation}

\subsection{Helmholtz free energy}
\label{subsec:free-energy}

Since the present black hole is uncharged and no cosmological
pressure is introduced, the appropriate thermodynamic potential is
the Helmholtz free energy,
\begin{equation}
 F_{\rm EGUP}
 =
 M-T_{\rm EGUP}S_{\rm EGUP}.
 \label{eq:helmholtz-definition}
\end{equation}
Using the mass--horizon relation, this becomes
\begin{equation}
 F_{\rm EGUP}(r_+)
 =
 \frac{3\pi}{8G_5}
 \left(
 r_+^2+\alpha_{\rm GB}
 \right)
 -
 T_{\rm EGUP}(r_+)S_{\rm EGUP}(r_+).
 \label{eq:helmholtz-explicit}
\end{equation}
The entropy normalization $S_{\rm ref}$ shifts the free energy by
$-T_{\rm EGUP}S_{\rm ref}$. Therefore, the reference state must be
specified before the absolute sign of the free energy is interpreted.

The free-energy curves below describe variations along a branch and do
not establish equilibrium preference between states at different temperatures.

\section{Critical behavior and endpoint structure}
\label{sec:critical-behavior}

To analyze the thermodynamic behavior independently of the choice of
units, we introduce the dimensionless quantities
\begin{equation}
 \rho=\frac{r_+}{\ell_{P,5}},
 \qquad
 a=\frac{\alpha_{\rm GB}}{\ell_{P,5}^2},
 \qquad
 \bar{\delta}=\delta\ell_{P,5}^2
 =\frac{\pi^2\eta^2\ell_{P,5}^2}{L^2}.
 \label{eq:dimensionless-parameters}
\end{equation}
The dimensionless effective localization scale and temperature are
defined by
\begin{equation}
 x(\rho)=\frac{X(r_+)}{\ell_{P,5}}
 =\rho+\frac{2a}{\rho},
 \qquad
 \tau=\ell_{P,5}T_{\rm EGUP}.
 \label{eq:dimensionless-X-temperature}
\end{equation}
Equation~\eqref{eq:temperature-rationalized} then becomes
\begin{equation}
 {
 \tau(\rho)
 =
 \frac{\pi x}{\beta^2}
 \left[
 1-\clearsqrt{
 1-\frac{\beta^2(1+\bar{\delta}x^2)}
 {\pi^2x^2}
 }
 \right]
 }.
 \label{eq:dimensionless-temperature}
\end{equation}
Equivalently,
\begin{equation}
 \tau(\rho)
 =
 \frac{1+\bar{\delta}x^2}
 {\pi x
 \left[
 1+\clearsqrt{
 1-\dfrac{\beta^2(1+\bar{\delta}x^2)}
 {\pi^2x^2}
 }
 \right]}.
 \label{eq:dimensionless-temperature-rational}
\end{equation}

The dimensionless mass can be written as
\begin{equation}
 m=
 \frac{8G_5M}{3\pi\ell_{P,5}^2}
 =\rho^2+a.
 \label{eq:dimensionless-mass}
\end{equation}
Thus, the mass is a monotonically increasing function of the horizon
radius, even though $x(\rho)$ is not monotonic.

The nonmonotonic localization scale also determines the small-radius
temperature behavior. It obeys the exact inversion symmetry
\begin{equation}
 x(\rho)=x\!\left(\frac{2a}{\rho}\right),
 \qquad \tau(\rho)=\tau\!\left(\frac{2a}{\rho}\right),
 \label{eq:inversion-symmetry}
\end{equation}
for $a>0$. Hence a temperature minimum at a large radius has a
small-radius partner with the same temperature. For $\bar\delta>0$
and $\beta^2\bar\delta<\pi^2$, the small-radius asymptote is
\begin{equation}
 \tau(\rho)\sim\frac{A}{\rho},\qquad
 A=\frac{2a\bar\delta}
 {\pi\left(1+\clearsqrt{1-\beta^2\bar\delta/\pi^2}\right)}>0.
 \label{eq:small-radius-temperature}
\end{equation}
Thus $d\tau/d\rho<0$ and $C_{\rm EGUP}<0$ sufficiently close to zero.
This is a formal asymptote of the phenomenological model, rather than a
controlled microscopic prediction at sub-Planckian radii.

Writing $K=1-\beta^2\bar\delta/\pi^2$, the stationary condition
$d\tau/dx=0$ gives
\begin{equation}
 x_*=\frac{1}{\clearsqrt{\bar\delta K}},\qquad
 \tau_{\min}=\frac{\clearsqrt{\bar\delta}}{\pi\sqrt K}.
 \label{eq:paired-minimum-scale}
\end{equation}
When $x_*^2>8a$, the two corresponding radii are
\begin{equation}
 \rho_{\min,s/l}=\frac{x_*\mp\sqrt{x_*^2-8a}}{2},
 \qquad \rho_{\min,s}\rho_{\min,l}=2a.
 \label{eq:paired-minimum-radii}
\end{equation}
These conditions hold for the numerical examples below. The coincident
case $x_*^2=8a$ requires a separate stationary-point classification.

\subsection{Admissible horizon-radius domain}
\label{subsec:admissible-domain}

The reality condition for the temperature takes the form
\begin{equation}
 \left(\pi^2-\beta^2\bar{\delta}\right)x^2
 \geq\beta^2,
 \qquad
 \pi^2>\beta^2\bar{\delta}.
 \label{eq:dimensionless-reality}
\end{equation}
Defining
\begin{equation}
 x_c=
 \frac{\beta}
 {\clearsqrt{\pi^2-\beta^2\bar{\delta}}},
 \label{eq:dimensionless-xc}
\end{equation}
the boundary radii are
\begin{equation}
 \rho_c^\pm=
 \frac{1}{2}
 \left(
 x_c\pm\sqrt{x_c^2-8a}
 \right),
 \label{eq:dimensionless-critical-radii}
\end{equation}
provided that
\begin{equation}
 x_c^2\geq8a.
 \label{eq:branch-existence-condition}
\end{equation}
The corresponding dimensionless masses are
\begin{equation}
 m_c^\pm=a+(\rho_c^\pm)^2.
 \label{eq:dimensionless-critical-masses}
\end{equation}

When the inequality in
Eq.~\eqref{eq:branch-existence-condition} is strict, the admissible
domain consists of
\begin{equation}
 0<\rho\leq\rho_c^-,
 \qquad\text{and}\qquad
 \rho\geq\rho_c^+.
 \label{eq:dimensionless-admissible-branches}
\end{equation}
The interval $\rho_c^-<\rho<\rho_c^+$ is excluded. An evaporating
black hole that begins on the large-radius branch cannot continuously
cross this interval and reach the small-radius branch.

At either boundary, the temperature remains finite and is given by
\begin{equation}
 \tau_c=
 \frac{\pi x_c}{\beta^2}.
 \label{eq:dimensionless-boundary-temperature}
\end{equation}

\subsection{Heat capacity at the branch endpoints}
\label{subsec:endpoint-capacity}

For the generic case $x_c^2>8a$, one has
\begin{equation}
 x'(\rho_c^+)>0,
 \qquad
 x'(\rho_c^-)<0.
\end{equation}
Since $Z\to0$ at both boundary radii, the derivative of the
temperature becomes singular, while the heat capacity approaches
zero. More precisely,
\begin{equation}
 \lim_{\rho\to(\rho_c^+)^+}
 C_{\rm EGUP}=0^-,
 \label{eq:outer-endpoint-capacity}
\end{equation}
whereas
\begin{equation}
 \lim_{\rho\to(\rho_c^-)^-}
 C_{\rm EGUP}=0^+.
 \label{eq:inner-endpoint-capacity}
\end{equation}
Therefore, the endpoint of the large-radius evaporation branch has
vanishing heat capacity but is approached from the locally unstable
side. A finite limiting temperature together with
$C_{\rm EGUP}\to0$ does not by itself establish a dynamically stable
remnant.

The degenerate case
\begin{equation}
  x_c^2=8a   
\end{equation}
requires separate treatment because $Z$ and $x'(\rho)$ vanish
simultaneously at
\begin{equation}
 \rho_c=\sqrt{2a}.
\end{equation}
Near the degenerate radius, the square-root function behaves as
$Z\propto|\rho-\rho_c|$.  The temperature therefore develops a cusp
rather than a differentiable stationary point.  The one-sided heat
capacities are finite and satisfy
\begin{equation}
 \lim_{\rho\to\rho_c^-}C_{\mathrm{EGUP}}
 =
 +\frac{3\pi}{2}\beta_{\mathrm{cr}}a,
 \qquad
 \lim_{\rho\to\rho_c^+}C_{\mathrm{EGUP}}
 =
 -\frac{3\pi}{2}\beta_{\mathrm{cr}}a.
\end{equation}
Thus, unlike the generic two-boundary case, the heat capacity neither
vanishes nor diverges at the degenerate radius.  Its sign changes
discontinuously across the temperature cusp.

\subsection{Large-radius behavior}
\label{subsec:large-radius}

For $\rho^2\gg 2a$, one has $x\simeq\rho$. In the presence of a
nonzero EUP deformation, the leading behavior of the temperature is
\begin{equation}
 \tau(\rho)
 \simeq
 \frac{\bar{\delta}\rho}
 {\pi\left[
 1+\clearsqrt{1-\beta^2\bar{\delta}/\pi^2}
 \right]}.
 \label{eq:large-radius-temperature}
\end{equation}
Thus, the EGUP temperature grows linearly at sufficiently large
radius. This behavior differs qualitatively from the GUP and
semiclassical limits, for which
\begin{equation}
 \tau_{\rm GUP}\sim\frac{1}{2\pi\rho},
 \qquad
 \tau_H\sim\frac{1}{2\pi\rho}.
\end{equation}

It follows that the heat capacity is positive in the asymptotic
large-radius region when $\bar{\delta}>0$. Since the heat capacity
near the outer limiting radius is negative, the outer branch is
expected to contain at least one divergence of $C_{\rm EGUP}$,
corresponding to an extremum of the temperature. The precise
location of this point is determined by
\begin{equation}
 \frac{d\tau}{d\rho}=0{,}
\end{equation}
{and is given by Eqs.~\eqref{eq:paired-minimum-scale} and
\eqref{eq:paired-minimum-radii} when the paired minima exist.}

\subsection{Thermodynamic interpretation}
\label{subsec:thermodynamic-interpretation}

The endpoint structure must be distinguished from local
thermodynamic stability. We shall use the following terminology:
\begin{itemize}
 \item a \emph{limiting configuration} is a boundary of the
 admissible EGUP domain;
 \item a \emph{locally stable configuration} satisfies
 $C_{\rm EGUP}>0$;
 \item a \emph{candidate thermodynamic remnant} is a limiting
 configuration with vanishing heat capacity;
 \item a {\emph{dynamically viable remnant}
 additionally requires stability under gravitational perturbations
 and a sufficiently suppressed evaporation rate.}
\end{itemize}

{Local thermodynamic stability should also be
distinguished from dynamical stability under gravitational
perturbations. In five-dimensional {EGB} gravity,
sufficiently large values of the {GB} coupling relative to
the horizon scale can trigger a high-multipole gravitational
instability \cite{Konoplya:2008ix}. Therefore, a positive heat
capacity alone does not guarantee the dynamical viability of a
configuration.}

The present analysis identifies limiting configurations and their
local thermodynamic properties. Whether any of these endpoints
represents a genuine remnant requires an analysis of the evaporation
rate{, including the relevant greybody factors,} and
cannot be concluded from the uncertainty relation and heat capacity
alone. {Indeed, {GB} corrections have been
shown to suppress graviton emission substantially and thereby
increase the lifetime of higher-dimensional black holes within the
dynamically stable parameter regime \cite{Konoplya:2010vz}.}

\section{Numerical results and discussion}
\label{sec:numerical}

We now examine the thermodynamic structure numerically using the
dimensionless variables introduced in the preceding section.  Unless stated
otherwise, we set
\begin{equation}
 a=0.01,
 \qquad
 \bar{\delta}=0.10,
\end{equation}
and vary the GUP parameter $\beta$.  These values satisfy the consistency
condition $\beta^{2}\bar{\delta}<\pi^{2}$ throughout the parameter ranges
considered below.

Three representative values of $\beta$ are selected:
\begin{equation}
 \beta=0.50,
 \qquad
 \beta=\beta_{\mathrm{cr}}=0.885043,
 \qquad
 \beta=1.20.
\end{equation}
They correspond, respectively, to the single-domain, degenerate, and
two-branch sectors of the parameter space.

\subsection{Parameter-space structure}
\label{subsec:parameter-space}

\begin{figure}[ht]
 \centering
 \includegraphics[width=0.65\textwidth]
 {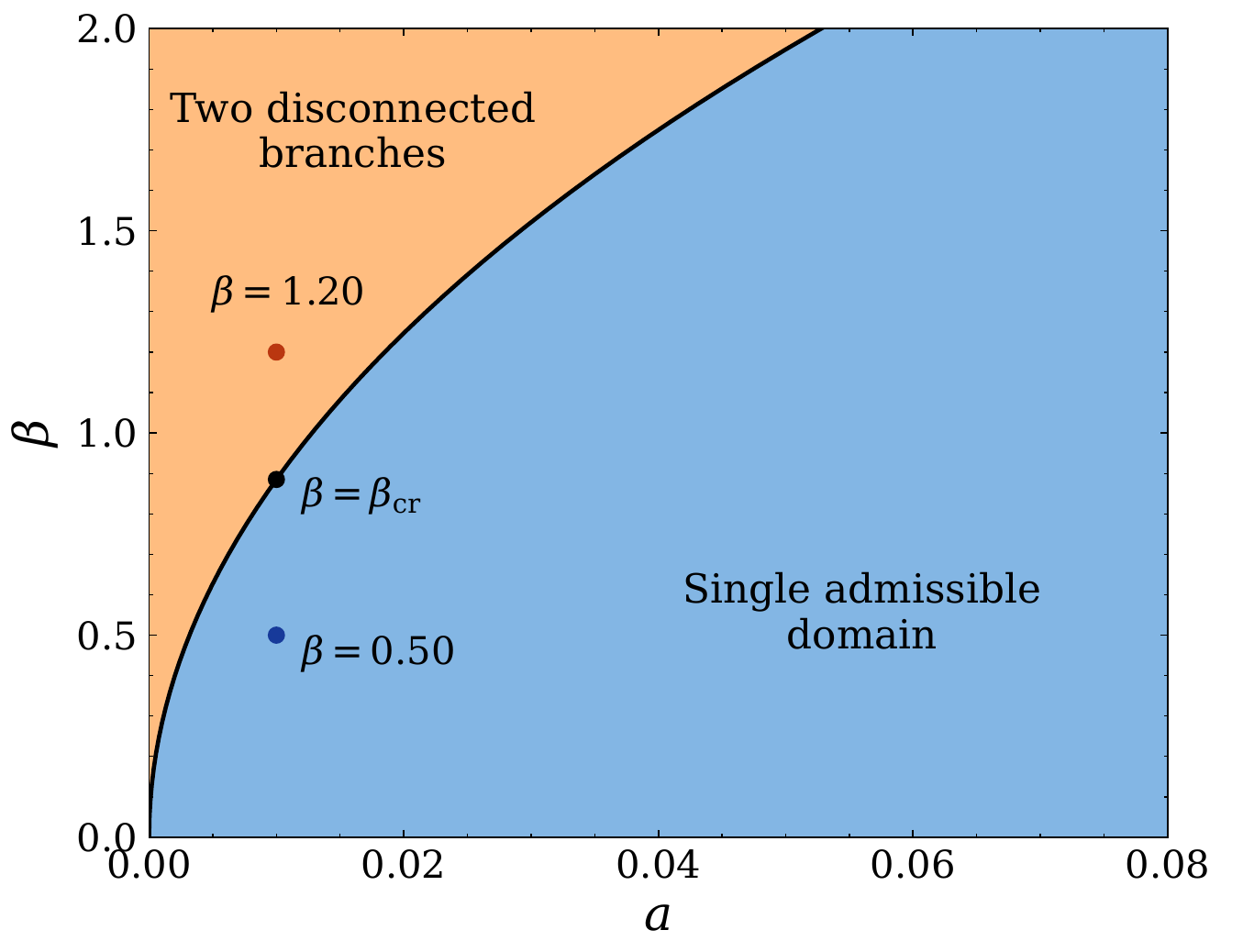}
 \caption{Parameter-space structure in the $(a,\beta)$ plane for
 fixed $\bar{\delta}=0.10$.  The solid curve represents
 $\beta=\beta_{\mathrm{cr}}$.  Below the curve the temperature is real
 throughout the positive-radius domain, whereas above it the admissible
 domain separates into two disconnected branches.  The three marked points
 indicate the representative parameter choices used in the numerical
 analysis.}
 \label{fig:parameter-space}
\end{figure}

Figure~\ref{fig:parameter-space} shows the parameter-space division in the
$(a,\beta)$ plane for fixed $\bar{\delta}=0.10$.  The separating curve is
\begin{equation}
 \beta_{\mathrm{cr}}(a,\bar{\delta})
 =
 \left(
 \frac{8\pi^{2}a}{1+8a\bar{\delta}}
 \right)^{1/2}.
\end{equation}
Below this curve, the EGUP temperature is real for every positive horizon
radius.  Above it, the admissible domain separates into two disconnected
branches.  On the curve itself, the two boundary radii coincide at
$\rho=\sqrt{2a}$.

For $a=0.01$, the critical value is
$\beta_{\mathrm{cr}}=0.885043$.  The corresponding critical quantities are
listed in Table~\ref{tab:critical-quantities}.  In the degenerate case,
\begin{equation}
 \rho_c^-=\rho_c^+=0.1414214,
 \qquad
 \tau_c=1.134399,
 \qquad
 m_c=0.0300.
\end{equation}
For $\beta=1.20$, the admissible branches are bounded by
\begin{equation}
 \rho_c^-=0.0619504,
 \qquad
 \rho_c^+=0.3228389.
\end{equation}
Both boundary configurations have the same finite temperature,
$\tau_c=0.839480$, but correspond to different masses,
$m_c^-=0.0138379$ and $m_c^+=0.114225$.

\begin{table}[ht]
 \centering
 \caption{Critical quantities for $a=0.01$ and
 $\bar{\delta}=0.10$.  A dash indicates that no EGUP-induced limiting
 radius occurs.}
 \label{tab:critical-quantities}
 \begin{tabular}{c c c c c c c}
  \hline\hline
  $\beta$
  & Region
  & $\rho_c^-$
  & $\rho_c^+$
  & $\tau_c$
  & $m_c^-$
  & $m_c^+$
  \\
  \hline
  $0.50$
  & Single domain
  & --
  & --
  & --
  & --
  & --
  \\
  $0.885043$
  & Degenerate
  & $0.1414214$
  & $0.1414214$
  & $1.134399$
  & $0.0300$
  & $0.0300$
  \\
  $1.20$
  & Two branches
  & $0.0619504$
  & $0.3228389$
  & $0.839480$
  & $0.0138379$
  & $0.114225$
  \\
  \hline\hline
 \end{tabular}
\end{table}

The two limiting masses in the last row should not be interpreted as two
successive endpoints of a single evaporation process.  A black hole
evaporating along the large-radius branch reaches $\rho_c^+$ and cannot
continuously cross the excluded interval to the small-radius branch.

\subsection{Temperature profiles}
\label{subsec:temperature-profiles}

\begin{figure*}[ht]
 \centering
 \includegraphics[width=0.96\textwidth]
 {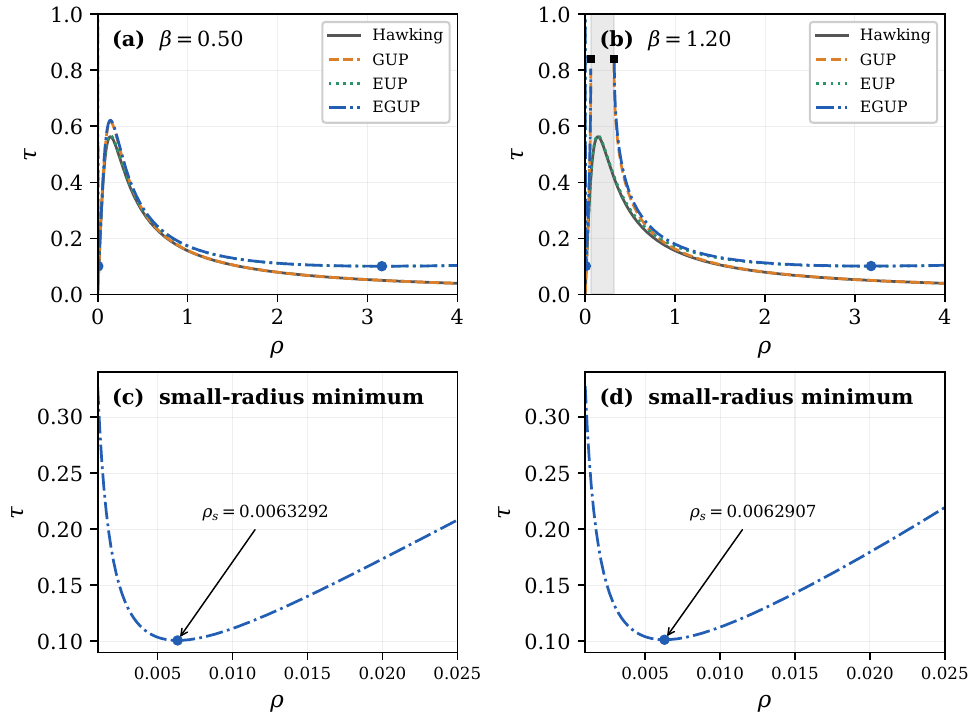}
 \caption{{Comparison of the dimensionless Hawking, GUP,
EUP, and EGUP temperatures for $a=0.01$ and $\bar\delta=0.10$.
Panels (a) and (b) show the profiles for $\beta=0.50$ and $1.20$;
panels (c) and (d) resolve their small-radius EGUP minima. All radius
axes are linear. Blue circles mark the EGUP minima, black squares mark
the branch endpoints, and shading denotes the interval excluded for
EGUP. Each model is drawn only where its temperature is real.}}
 \label{fig:temperature-comparison}
\end{figure*}

We first compare the semiclassical, GUP, EUP, and full EGUP temperature
prescriptions.  The results are displayed in
Fig.~\ref{fig:temperature-comparison} for the representative single-domain
and two-branch parameter choices.

In the single-domain case, the full EGUP temperature remains real for every
$\rho>0$.  It reaches a local maximum at
\begin{equation}
 \rho_{\mathrm{max}}=0.14142136,
 \qquad
 \tau_{\mathrm{max}}=0.62154423.
\end{equation}
The position of this maximum coincides with the minimum of the effective
localization scale,
\begin{equation}
 \rho_{\mathrm{max}}=\sqrt{2a}.
\end{equation}
This follows because the temperature depends on the horizon radius through
$x(\rho)$ and $x'(\rho)$ vanishes at this point.

The same single-domain profile also has a small-radius minimum at
\begin{equation}
 \rho_{\min,s}=0.006329192,\qquad \tau_{\min,s}=0.10078615.
 \label{eq:numerical-small-minimum-connected}
\end{equation}
The temperature decreases from its inverse-radius divergence to this
minimum, rises to the local maximum at $\sqrt{2a}$, and then decreases
toward the large-radius minimum. The small-radius behavior is resolved
in the small-radius panels of Fig.~\ref{fig:temperature-comparison}.

At larger radius, the competition between the decreasing Hawking-like term
and the increasing EUP contribution {produces the large-radius temperature minimum.}
For $\beta=0.50$, this minimum occurs at
\begin{equation}
 \rho_{\mathrm{min}}=3.1599612,
 \qquad
 \tau_{\mathrm{min}}=0.10078615.
\end{equation}
Beyond this point, the temperature increases rather than approaching zero.

For $\beta=1.20$, the stationary point at $\rho=\sqrt{2a}$ lies inside the
excluded interval and therefore does not belong to either physical branch.
The small-radius branch terminates at $\rho_c^-$, while the large-radius
branch begins at $\rho_c^+$.  Both approach the same finite boundary
temperature, although they are not continuously connected.  The
temperature decreases immediately outside $\rho_c^+$ and reaches its
large-radius minimum at
\begin{equation}
 \rho_{\mathrm{min}}=3.1793118,
 \qquad
 \tau_{\mathrm{min}}=0.10140087.
\end{equation}

Its small-radius partner lies on the admissible small branch:
\begin{equation}
 \rho_{\min,s}=0.006290670,\qquad \tau_{\min,s}=0.10140087,
 \label{eq:numerical-small-minimum-disconnected}
\end{equation}
which satisfies $\rho_{\min,s}<\rho_c^-$. The two minima have equal
temperatures by Eq.~\eqref{eq:inversion-symmetry}.

\begin{figure*}[ht]
 \centering
 \includegraphics[width=0.96\textwidth]
 {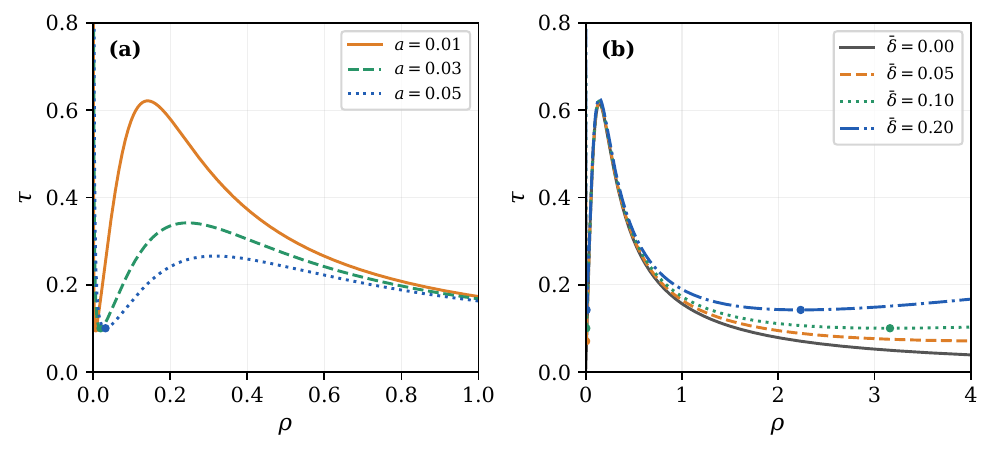}
 \caption{{Dependence of the exact EGUP temperature on the deformation
parameters. Panel (a) varies $a$ at $\beta=0.50$, $\bar\delta=0.10$;
panel (b) varies $\bar\delta$ at $a=0.01$, $\beta=0.50$.
Linear radius axes are used. Circles mark temperature minima within
the plotted ranges; the small-radius minima are resolved in
Fig.~\ref{fig:temperature-comparison}. For
$\bar\delta=0$ the small-radius temperature tends to zero instead.}}
 \label{fig:temperature-parameters}
\end{figure*}

The separate roles of the {GB} and EUP parameters are illustrated
in Fig.~\ref{fig:temperature-parameters}.  Increasing $a$ moves the
temperature maximum to larger radius because
$\rho_{\mathrm{max}}=\sqrt{2a}$.  At the same time, the maximum temperature
decreases.  Thus, the {GB} scale both shifts and suppresses the
near-horizon temperature peak.

The large-radius behavior is controlled by $\bar{\delta}$.  For
$\bar{\delta}=0$, the temperature decreases asymptotically as
$\tau\sim(2\pi\rho)^{-1}$.  When $\bar{\delta}>0$, the temperature instead
reaches a nonzero minimum and subsequently grows linearly.  In particular,
for $\rho^2\gg 2a$,
\begin{equation}
 \tau(\rho)
 \simeq
 \frac{\bar{\delta}\rho}
 {\pi\left(1+\clearsqrt{1-\beta^{2}\bar{\delta}/\pi^{2}}\right)}.
\end{equation}
Increasing $\bar{\delta}$ raises both the position-independent correction
at intermediate radii and the asymptotic slope.  This behavior is absent
in the pure GUP and semiclassical limits and is entirely associated with
the minimum-momentum sector.

\subsection{Entropy behavior}
\label{subsec:numerical-entropy}

The exact EGUP entropy is evaluated numerically from the first law.  Since
the entropy is defined only up to an additive constant, we work with the
difference
\begin{equation}
 \Delta S_{\mathrm{EGUP}}(\rho;\rho_{\mathrm{ref}})
 =
 S_{\mathrm{EGUP}}(\rho)
 -
 S_{\mathrm{EGUP}}(\rho_{\mathrm{ref}}).
\end{equation}
For the connected single-domain configuration, we choose
$\rho_{\mathrm{ref}}=1$.

\begin{figure}[ht]
 \centering
 \includegraphics[width=0.65\textwidth]
 {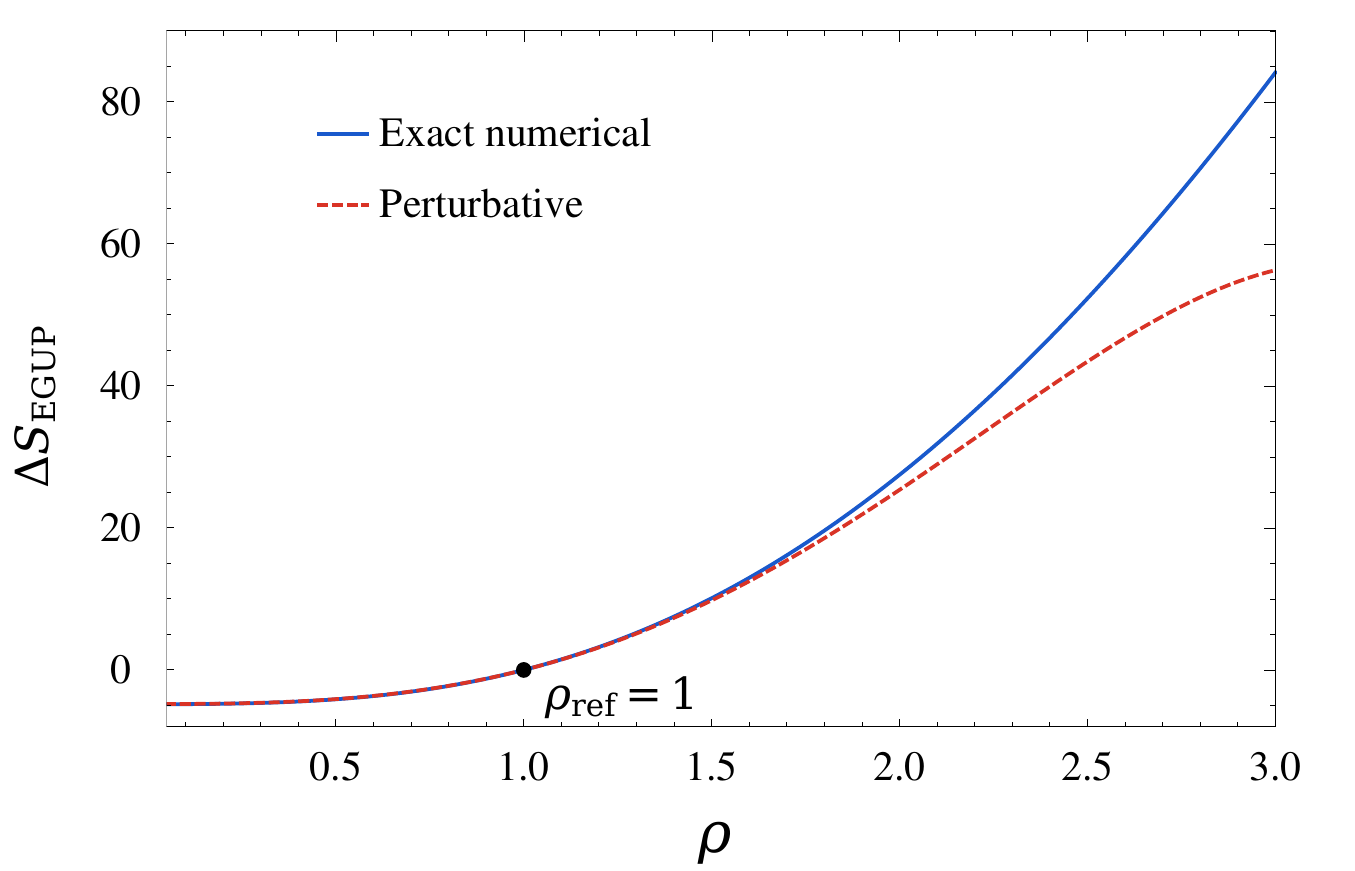}
 \caption{Exact numerical and perturbative EGUP entropy differences for
 $a=0.01$, $\beta=0.50$, $\bar{\delta}=0.10$, and
 $\rho_{\mathrm{ref}}=1$.  The black point denotes the reference
 configuration at which both entropy differences vanish.}
 \label{fig:entropy-comparison}
\end{figure}

Figure~\ref{fig:entropy-comparison} compares the exact numerical result with
the perturbative expression derived previously.  Both results vanish at
the reference radius by construction.  They remain close in the vicinity
of $\rho_{\mathrm{ref}}$, but gradually separate as the horizon radius
increases.

At $\rho=3$, the two results are
\begin{equation}
 \Delta S_{\mathrm{EGUP}}^{\mathrm{exact}}
 =
 84.1206,
 \qquad
 \Delta S_{\mathrm{EGUP}}^{\mathrm{pert}}
 =
 56.2830.
\end{equation}
Thus, for the parameters used here, the truncated expansion underestimates
the exact entropy difference by approximately $33\%$ at $\rho=3$.  This
departure is expected because the perturbative expansion requires not only
small $\beta$ and $\bar{\delta}$ separately, but also
$\bar{\delta}x^{2}\ll1$.  
This condition can fail at both small and large radii, as explained
after Eq.~\eqref{eq:entropy-validity}. The exact entropy integral must
therefore be used outside the perturbative window, including either
radial extreme.

In the two-branch sector, the entropy must be normalized independently on
each connected branch.  For the small-radius branch, choosing
$\rho_{\mathrm{ref}}^-=0.01$ gives
\begin{equation}
 \Delta S_{\mathrm{EGUP}}
 \bigl(\rho_c^-;\rho_{\mathrm{ref}}^-\bigr)
 \simeq
 0.01305.
\end{equation}
For the large-radius branch, choosing the outer limiting configuration as
the reference gives
\begin{equation}
 \Delta S_{\mathrm{EGUP}}
 \bigl(3;\rho_c^+\bigr)
 \simeq
 87.49.
\end{equation}
In both cases the entropy increases monotonically along the corresponding
branch because $dM/d\rho>0$ and the physical temperature is positive.

The additive constants on the two disconnected branches cannot be fixed
relative to one another using the first law alone.  Consequently, absolute
entropy values on the small- and large-radius branches should not be
compared without an additional prescription for the reference states.

\subsection{Heat capacity and stability regions}
\label{subsec:numerical-heat-capacity}

\begin{figure*}[ht]
 \centering
 \includegraphics[width=0.98\textwidth]
 {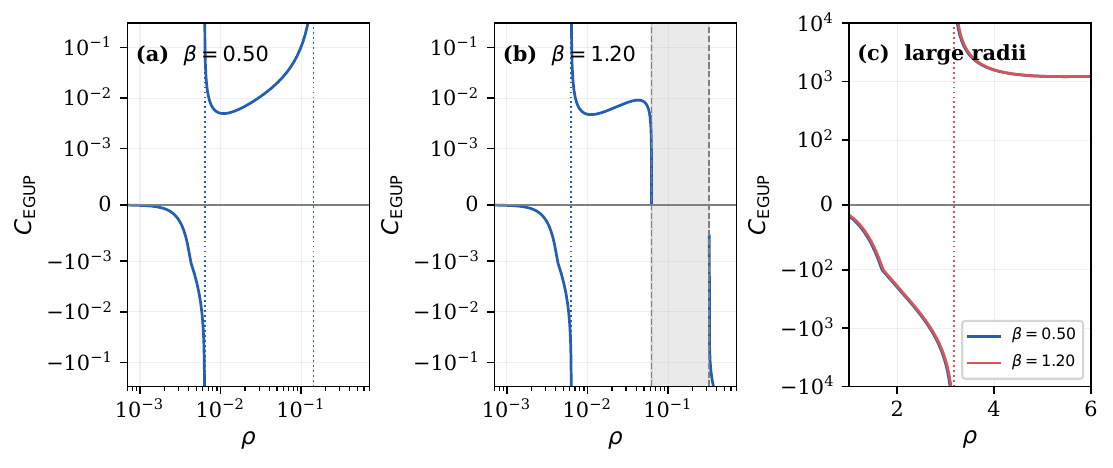}
 \caption{{EGUP heat capacity for $a=0.01$ and $\bar\delta=0.10$.
Panels (a) and (b) use logarithmic radius axes and symmetric logarithmic
heat-capacity axes to resolve the small-radius instability and the
additional $-\to+$ divergence for $\beta=0.50$ and $1.20$, respectively.
Panel (c) compares the large-radius minima. Dotted vertical lines mark
temperature extrema; shading and dashed boundaries identify the excluded
interval for $\beta=1.20$. Curves are split at all poles and excluded boundaries.}}
 \label{fig:heat-capacity}
\end{figure*}

The local thermodynamic stability is determined by the sign of
$C_{\mathrm{EGUP}}$.  The numerical results are shown in
Fig.~\ref{fig:heat-capacity}.  Because the heat capacity varies rapidly
near its divergences and the small- and large-radius sectors have very
different scales, separate panels are used for the near-horizon and
large-radius regimes.

For $\beta=0.50$, the correct sign structure is
\begin{equation}
\begin{aligned}
 C_{\rm EGUP}&<0, &&0<\rho<0.006329192,\\
 C_{\rm EGUP}&>0, &&0.006329192<\rho<0.14142136,\\
 C_{\rm EGUP}&<0, &&0.14142136<\rho<3.1599612,\\
 C_{\rm EGUP}&>0, &&\rho>3.1599612.
\end{aligned}
 \label{eq:corrected-connected-stability}
\end{equation}
The small-radius minimum gives a divergence with a $-\to+$ sign change;
the maximum at $\sqrt{2a}$ gives a $+\to-$ change; and the large-radius
minimum gives the final $-\to+$ change. Positive heat capacity therefore
does not extend to arbitrarily small radii.

For $\beta=1.20$, the small branch instead satisfies
\begin{equation}
\begin{aligned}
 C_{\rm EGUP}&<0, &&0<\rho<0.006290670,\\
 C_{\rm EGUP}&>0, &&0.006290670<\rho<\rho_c^-.
\end{aligned}
 \label{eq:corrected-disconnected-stability}
\end{equation}
It approaches its endpoint with $C_{\rm EGUP}\to0^+$. The large branch
begins with $C_{\rm EGUP}\to0^-$ at $(\rho_c^+)^+$, remains locally
unstable up to $\rho_{\min,l}=3.1793118$, and has $C_{\rm EGUP}>0$
thereafter. The forbidden interval separates the two branches and is
not crossed by a continuous evaporation trajectory.

The characteristic temperature extrema and stability changes are
summarized in Table~\ref{tab:temperature-extrema}.

\begin{table*}[ht]
\centering
\caption{{Temperature extrema and heat-capacity sign changes
for $a=0.01$ and $\bar\delta=0.10$, including both small-radius minima.}}
\label{tab:temperature-extrema}
\setlength{\tabcolsep}{4pt}
\begin{tabular}{cccccc}
\hline\hline
$\beta$ & Branch & Extremum & $\rho_{\rm ext}$ & $\tau_{\rm ext}$ & Change in $C$\\
\hline
$0.50$ & Connected & Small minimum & $0.006329192$ & $0.10078615$ & $-\to+$\\
$0.50$ & Connected & Maximum & $0.14142136$ & $0.62154423$ & $+\to-$\\
$0.50$ & Connected & Large minimum & $3.1599612$ & $0.10078615$ & $-\to+$\\
$1.20$ & Small-radius & Minimum & $0.006290670$ & $0.10140087$ & $-\to+$\\
$1.20$ & Large-radius & Minimum & $3.1793118$ & $0.10140087$ & $-\to+$\\
\hline\hline
\end{tabular}
\end{table*}

At the critical value $\beta=\beta_{\mathrm{cr}}=0.885043$, the
temperature has a cusp at $\rho_c=0.1414214$. The corresponding
one-sided heat capacities are
\begin{equation}
 \lim_{\rho\to\rho_c^-}C_{\mathrm{EGUP}}
 \simeq +0.04171,
 \qquad
 \lim_{\rho\to\rho_c^+}C_{\mathrm{EGUP}}
 \simeq -0.04171,
\end{equation}
in agreement with the analytic result obtained above. The degenerate
critical radius therefore separates locally stable and unstable sectors
without producing a divergence or a vanishing heat capacity.

The vanishing heat capacity at $\rho_c^+$ identifies the outer boundary as
a candidate thermodynamic endpoint.  Nevertheless, it is approached from
the locally unstable side and has a finite temperature.  These properties
do not establish a dynamically stable remnant.  Such a conclusion would
require an independent calculation of the evaporation rate and its
behavior near the limiting configuration.

Similarly, the divergences of $C_{\mathrm{EGUP}}$ identify continuous
stability changes within the present thermodynamic description.  They
should not by themselves be interpreted as evidence for a global phase
transition, particularly because the underlying spacetime is
asymptotically flat.

\subsection{Free-energy behavior}
\label{subsec:numerical-free-energy}

To describe the variation of the free energy along each connected branch,
we introduce the dimensionless Helmholtz free energy

\begin{equation}
 \mathcal{F}
 =
 \frac{8G_5F}{3\pi\ell_{P,5}^{\,2}}
 =
 m-\frac{8}{3\pi}\tau S_{\mathrm{EGUP}}.
\end{equation}

\begin{figure*}[ht]
 \centering
 \includegraphics[width=0.96\textwidth]
 {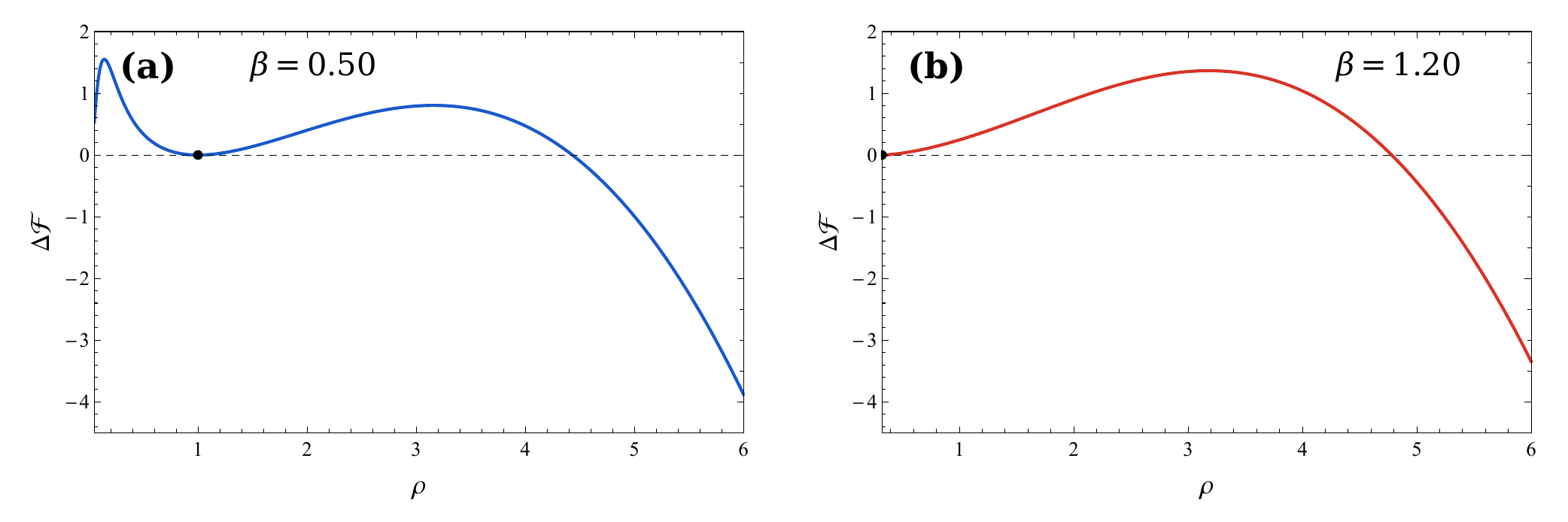}
 \caption{{Dimensionless Helmholtz free-energy differences for
$a=0.01$ and $\bar\delta=0.10$. The entropy is normalized by
$S_{\rm EGUP}(\rho_{\rm ref})=0$ in each panel: (a) $\beta=0.50$,
$\rho_{\rm ref}=1$; (b) the large-radius branch for $\beta=1.20$,
$\rho_{\rm ref}=\rho_c^+$. Dashed horizontal lines mark
$\Delta\mathcal F=0$. The curves are parametric variations along
the branches, not comparisons of equilibrium preference at a common temperature.}}
 \label{fig:free-energy}
\end{figure*}

For a specified entropy normalization we plot the difference
\begin{equation}
 \Delta\mathcal F(\rho;\rho_{\rm ref})
 =\mathcal F(\rho)-\mathcal F(\rho_{\rm ref}).
 \label{eq:free-energy-difference}
\end{equation}
The normalization used in Fig.~\ref{fig:free-energy} is explicitly
\begin{equation}
 S_{\rm EGUP}(\rho)=\int_{\rho_{\rm ref}}^{\rho}
 \frac{3\pi u}{4\tau(u)}\,du,\qquad S_{\rm EGUP}(\rho_{\rm ref})=0.
 \label{eq:figure-entropy-normalization}
\end{equation}
Here $G_5=\ell_{P,5}^3$ has been used. Thus $\mathcal F(\rho_{\rm ref})
=m(\rho_{\rm ref})$. Subtracting this reference value does not remove
the entropy-normalization dependence: shifting $S$ by a constant $s_0$
changes $\Delta\mathcal F$ by
$-8s_0[\tau(\rho)-\tau(\rho_{\rm ref})]/(3\pi)$.

For the single-domain case we use $\rho_{\mathrm{ref}}=1$.  In the
two-branch case, we restrict the analysis to the large-radius branch and
choose $\rho_{\mathrm{ref}}=\rho_c^+$.  The resulting free-energy
differences are displayed in Fig.~\ref{fig:free-energy}.

In the single-domain case, the stationary point at $\rho=1$ follows
from $S_{\rm EGUP}(1)=0$, not from the subtraction
$\Delta\mathcal F(1;1)=0$ alone. Indeed, at regular points the first law gives
\begin{equation}
 \frac{d\Delta\mathcal F}{d\rho}
 =-\frac{8}{3\pi}S_{\rm EGUP}(\rho)\frac{d\tau}{d\rho}.
 \label{eq:free-energy-derivative}
\end{equation}
At the outer branch endpoint in panel (b), the derivative of $\tau$
is singular, but $S=O(\rho-\rho_c^+)$ and
$\tau'=O((\rho-\rho_c^+)^{-1/2})$, so their product tends to zero
from the admissible side.

For $\beta=0.50$ and $\rho>1$, the free energy increases and reaches
its maximum at the large-radius temperature minimum,
$\rho=3.1599612$.
  It then decreases and crosses zero again at
\begin{equation}
 \rho\simeq4.43068.
\end{equation}
For the large-radius branch with $\beta=1.20$, the free-energy difference
starts from zero at $\rho_c^+$, {increases to a maximum at the temperature minimum at $\rho=3.1793118$}, and changes sign at
\begin{equation}
 \rho\simeq4.78026.
\end{equation}

The correspondence between the free-energy extrema and the temperature
extrema follows from the first law.  At fixed external parameters,
\begin{equation}
 dF=-S_{\mathrm{EGUP}}\,dT_{\mathrm{EGUP}}.
\end{equation}
Consequently, away from a reference point at which the normalized entropy
vanishes, an extremum of the temperature also produces an extremum of the
Helmholtz free energy.

The locations of the zero crossings depend on the entropy normalization.
They therefore do not represent invariant critical radii.  Moreover, since
the underlying spacetime is asymptotically flat and no thermodynamic
pressure or confining boundary is introduced, a sign change of
$\Delta\mathcal{F}$ should not be interpreted as a Hawking--Page phase
transition.  
The curves describe the variation of the free energy with radius under
the stated normalization. Points on them generally have different
temperatures, so their relative heights do not establish equilibrium
thermodynamic preference, even on the same connected branch. We do not
claim a comparison of competing saddles at a common bath temperature.

\section{Conclusion}
\label{sec:conclusion}

In this work, we have studied the thermodynamics of an asymptotically
flat five-dimensional {EGB} black hole within the
extended generalized uncertainty principle. Particular care was taken
to preserve dimensional consistency in five dimensions and to
distinguish the {GB} coupling in the action from the effective
coupling entering the black-hole metric.

Starting from the field equations, we obtained the physical black-hole
branch and its semiclassical thermodynamic quantities. The EGUP
temperature was then constructed using a geometry-dependent localization

scale calibrated to recover the surface-gravity temperature in the
undeformed limit. This phenomenological calibration and the GUP baseline
are inherited from Ref.~\cite{PradhanVargheseFairoos2026}; the present
work develops their EGUP extension and its admissible-domain structure.
 The resulting expression consistently reduces to the
GUP, EUP, and standard Hawking temperatures in the corresponding limits.
The simultaneous presence of the GUP and EUP sectors also requires
\begin{equation}
 \frac{\beta\eta\ell_{P,5}}{L}<1,
\end{equation}
which ensures that both the minimum-length and minimum-momentum scales
remain real.

A central result is that the EGUP reality condition does not generically
produce a unique minimum horizon radius. Depending on the {GB}
and deformation parameters, the temperature can remain real for all
positive radii, develop a degenerate critical radius, or admit two
disconnected radius branches separated by an excluded interval. In the
last case, a black hole evaporating along the large-radius branch reaches
the outer boundary and cannot continuously pass to the small-radius
branch.

The numerical analysis shows how the three relevant scales compete.
Increasing the {GB} coupling moves the near-horizon temperature
maximum to larger radius and reduces its height. The GUP contribution
mainly affects the short-distance regime and controls the appearance of
the disconnected branches. The EUP contribution instead governs the
large-radius behavior. For nonzero EUP deformation, the temperature
reaches a minimum and subsequently grows linearly, rather than vanishing
as in the semiclassical and pure-GUP cases.

For nonzero EUP deformation, the inversion symmetry of the localization
scale pairs the small- and large-radius temperature minima. The exact
temperature diverges as $\rho^{-1}$ near zero, so the heat capacity is
negative at sufficiently small radii. A locally stable interval begins
only after the small-radius minimum. In the connected example it ends
at the intervening maximum; a further stable interval begins after the
large-radius minimum. These local thermodynamic statements do not imply
gravitational stability or a controlled microscopic description near zero.
 In the two-branch case, the small branch approaches its
boundary with $C_{\mathrm{EGUP}}\to0^+$, whereas the accessible
large-radius branch begins with $C_{\mathrm{EGUP}}\to0^-$. The outer
limiting configuration therefore has a finite temperature and vanishing
heat capacity but is approached from the locally unstable side. These
properties alone are not sufficient to establish a dynamically stable
black-hole remnant.

The exact entropy obtained from the first law agrees with its
perturbative approximation near the chosen reference configuration, but

the two results separate as the EUP expansion parameter grows.
The perturbative entropy requires both conditions in
Eq.~\eqref{eq:entropy-validity} throughout the integration interval and
can fail at both small and large radii.
 We also find that the leading GUP correction to the entropy
survives in the Einstein-gravity limit $\alpha_{\rm GB}\to0$. This point
is relevant to the interpretation of earlier GUP treatments of the
five-dimensional {EGB} black hole.

Finally, the Helmholtz free energy describes variations along a branch
with the explicit normalization $S(\rho_{\rm ref})=0$. Its stationary
point at the regular reference radius follows from that normalization,
not from reference subtraction alone. Its zero crossings depend on the
entropy constant, and comparisons between different temperatures do not
establish equilibrium preference or a Hawking--Page transition.
 Determining whether the limiting
configurations found here can survive as genuine long-lived remnants
would require a direct calculation of the evaporation rate, including
the corresponding greybody factors and possible backreaction effects.
{{GB} corrections can substantially suppress
graviton emission and prolong black-hole lifetimes within the dynamically
stable regime \cite{Konoplya:2010vz}.}

\section*{Acknowledgments}

B. C. L. is grateful to the Excellence project FoS UHK 2205/2025--2026 for the financial support and Roman Konoplya for useful discussions.

\section*{Data Availability Statement}

No new data were created or analyzed in this study.

\section*{Conflict of Interest}

The authors declare no conflict of interest.

\bibliographystyle{unsrt}
\bibliography{references}

\end{document}